\documentclass[prx, twocolumn, notitlepage,superscriptaddress, nofootinbib]{revtex4-1}

\usepackage{fancyhdr}
\fancyheadoffset[L,R]{\headrulewidth}
\renewcommand{\headrulewidth}{0.6pt}

\usepackage{cancel}
\usepackage[normalem]{ulem}

\usepackage{cancel}

\usepackage{graphicx}
\usepackage{tikz,pgf}
\usepackage{color}
\usepackage{array}
\usepackage{xcolor}
\usepackage[dvips]{epsfig}
\usepackage{epsfig}
\usepackage{enumerate}
\usepackage[latin1]{inputenc}
\usepackage[T1]{fontenc}

\usepackage[hang]{subfigure}
\usepackage{bbm, dsfont}
\usepackage{amsfonts,amsmath,amssymb,stmaryrd}
\usepackage{ulem}

\usepackage{mathrsfs}
\usepackage{bbold}

\usepackage{simplewick}

\newcommand{\bra}[1]{\langle #1 | \,}
\newcommand{\ket}[1]{\, | #1 \rangle}

\newcommand{\ga}{\ga}

\newcommand{\bl}{\begin{linenomath*}}
\newcommand{\el}{\end{linenomath*}}

\newcommand{\bea}{\begin{eqnarray}}
\newcommand{\eea}{\end{eqnarray}}
\renewcommand{\vec}[1]{\mathbf{#1}}

\newcommand{\be}{\hat b}
\newcommand{\bed}{\hat b^\dagger}
\renewcommand{\ga}{\hat\gamma}

\newcommand{\bd}{\hat b^\dagger}

\usepackage{hyperref}

\begin{document}

\title{Deformation of a quantum many-particle system by a rotating impurity}

\author{Richard Schmidt} 
\email{richard.schmidt@cfa.harvard.edu}
\affiliation{ITAMP, Harvard-Smithsonian Center for Astrophysics, 60 Garden Street, Cambridge, MA 02138, USA}%
\affiliation{Physics Department, Harvard University, 17 Oxford Street, Cambridge, MA 02138, USA} %

\author{Mikhail Lemeshko} 
\email{mikhail.lemeshko@ist.ac.at}
\affiliation{IST Austria (Institute of Science and Technology Austria), Am Campus 1, 3400 Klosterneuburg, Austria}

\begin{abstract}

During the last 70 years, the quantum theory of angular momentum has been successfully applied to describing the properties of nuclei, atoms, and molecules, their interactions with each other as well as with external fields. Due to the properties of quantum rotations, the angular momentum algebra can be of tremendous complexity even for a few interacting particles, such as valence electrons of an atom, not to mention larger many-particle systems. In this work, we study an example of the latter: a rotating quantum impurity coupled to a many-body bosonic bath. In the regime of strong impurity-bath couplings the problem involves  addition of an  infinite number of angular momenta which renders it intractable using currently available techniques. Here, we introduce a novel canonical transformation which allows to eliminate the complex angular momentum algebra from such a class of many-body problems. In addition, the transformation exposes the problem's constants of motion, and renders it solvable exactly in the limit of a slowly-rotating impurity. We exemplify the technique by showing that there exists a critical rotational speed at which the impurity suddenly acquires one quantum of angular momentum from the many-particle bath.  Such an instability is accompanied by the deformation of the phonon density in the frame rotating along with the impurity.

\vspace{0.5cm}

\end{abstract}

\maketitle

\section{Introduction}

An important part of modern condensed matter physics deals with so-called `impurity problems', aiming to understand the behavior of individual quantum particles coupled to a complex many-body environment. The interest in quantum impurities goes back to the classic works of Landau, Pekar, Fr\"ohlich, and Feynman, who showed that propagation of electrons in crystals is largely affected by the quantum field of lattice excitations and can be rationalized by introducing the quasiparticle concept of the polaron~\cite{LandauPolaron, LandauPekarJETP48, FrohlichAdvPhys54, FeynmanPR55}. In turn, the properties of a quantum many-body system can be drastically modified by the presence of impurities. The most known examples are the Kondo effect~\cite{KondoPTP64} -- suppression of electron transport due to magnetic impurities in metals -- and the Anderson orthogonality catastrophe which leads to the edge singularities  in the X-ray absorption spectra of metals~\cite{AndersonPRL67}.

In many instances, the impurities -- even those possessing an internal structure -- can be accurately described as point-like particles. The latter is justified by the separation of the energy scales inherent to the impurity and the surrounding bath. A well-known example is that of Bose- and Fermi-polarons realized in cold atomic gases by a number of groups~\cite{ChikkaturPRL00, SchirotzekPRL09, PalzerPRL09, KohstallNature12, KoschorreckNature12, SpethmannPRL12, FukuharaNatPhys13, ScellePRL13, Cetina15, MassignanRPP14}.  There, the spherically symmetric ground state of an alkali atom lies hundreds of THz lower than any of its electronically excited states. Given ultracold collision energies, such an energy gap renders all the processes happening inside of an atom irrelevant.

More complex systems, such as molecules, are extended objects and therefore possess a number of fundamentally different types of internal motion. The latter stem from the relative motion of the nuclei, such as  rotation and vibration, which couple to each other as well as to the electronic spin and orbital degrees of freedom~\cite{LemKreDoyKais13, KreStwFrieColdMol, BernathBook, LevebvreBrionField2}. This results in a rich low-energy   dynamics which is highly susceptible to external perturbations. Moreover, in many experimental realizations molecular rotation is coupled to a phononic bath pertaining to the surrounding medium such as superfluid helium~\cite{ToenniesAngChem04}, rare-gas matrix~\cite{MatrixIsolationBook}, or a Coulomb crystal formed in an ion trap~\cite{WillitschIRPC12}, which needs to be properly accounted for by a microscopic theory.

The concept of orbital angular momentum, however, goes far beyond physically rotating systems and is being used to describe e.g.\ the excited-state electrons in solids, whose motion is perturbed by  lattice vibrations~\cite{Mahan90}, or Rydberg atoms immersed into a Bose-Einstein condensate~\cite{Greene2000,BalewskiNature13}. Despite the ubiquitous use of the angular momentum concept in various branches of physics, a versatile theory describing the redistribution of orbital angular momentum in quantum many-body systems has not yet been developed.

Recently,  we have undertaken the first step towards such a theory by deriving a generic Hamiltonian which describes the coupling of an $SO(3)$-symmetric impurity -- a quantum rotor -- with a bath of harmonic oscillators~\cite{SchmidtLem15}. We have shown that the problem can be approached most naturally by introducing the  quasiparticle concept of the  \textit{`angulon'} -- a quantum rotor dressed by a quantum field. The angulon is an eigenstate of the total angular momentum of the system, which remains a  conserved quantity in the presence of the impurity-bath interactions. It was found that even single-phonon excitations of the bath alone are capable of drastically modifying the rotational spectrum of the impurity, which manifests itself in the emerging Many-Body-Induced Fine Structure~\cite{SchmidtLem15}.

 Here we demonstrate that  rotation of an anisotropic impurity can, in turn, substantially alter the collective state of a many-particle system. The effects are most significant in the regime of strong correlations, which however requires adding an infinite number of angular momentum vectors pertaining to possible many-body states. The resulting angular momentum algebra involves Wigner $3nj$-symbols~\cite{VarshalovichAngMom} of an arbitrarily high order and is therefore intractable using standard techniques. In order to overcome this problem, here we introduce a canonical transformation, which, to our knowledge, has never appeared in the literature before.
The transformation renders the Hamiltonian independent of the impurity coordinates, thereby eliminating the complex angular momentum algebra from the many-body problem.  Furthermore, the transformation singles out the conserved quantities of the many-body problem and renders it solvable exactly in the limit of a slowly rotating impurity.
 
 The transformation makes it apparent that there exists a critical rotational speed which leads to an instability, accompanied by a discontinuity in the many-particle spectrum. Unlike in the vortex instability, originating from rotation of a condensate around a given axis~\cite{Pitaevskii2003}, the instability we uncover here corresponds to the finite transfer of three-dimensional angular momentum between the impurity and the bath. It exists  solely due to the discrete energy spectrum inherent to quantum rotation. We demonstrate that the emerging instability is ushered by a macroscopic deformation of the surrounding bath, i.e. the phonon density modulation in the frame co-rotating with the impurity.

 \section{The canonical transformation}

We start from the general Hamiltonian of the angulon problem, as defined in Ref.~\cite{SchmidtLem15}:
\begin{multline}
\label{Hamil}
\hat H= B \mathbf{\hat{J}^2} + \sum_{k \lambda \mu}  \omega_k \bed_{k\lambda \mu} \be_{k\lambda \mu} \\+   \sum_{k \lambda \mu} U_\lambda(k)  \left[ Y^\ast_{\lambda \mu} (\hat \theta,\hat \phi) \bed_{k \lambda \mu}+ Y_{\lambda \mu} (\hat \theta,\hat \phi) \be_{k \lambda \mu} \right],
\end{multline}
where $Y_{\lambda \mu} (\hat \theta,\hat \phi)$ are the spherical harmonics~\cite{VarshalovichAngMom} depending on the molecular angle \textit{operators} $\hat \theta$ and $\hat \phi$, $\sum_k\equiv\int dk$, and $\hbar \equiv 1$.

The first term of Eq.~(\ref{Hamil}) corresponds to the kinetic energy of the translationally-localized linear-rotor impurity, with $B$ the rotational constant and $\mathbf{\hat{J}}$ the angular momentum operator.  In the absence of an external bath, the impurity eigenstates, $\vert j, m \rangle$, are labeled by the angular momentum, $j$, and its projection, $m$,  onto the laboratory-frame $z$-axis. Unperturbed rotational states form $(2j+1)$-fold degenerate multiplets with energies $E_j = B j(j+1)$~\cite{LevebvreBrionField2, BernathBook, LemKreDoyKais13}.

\begin{figure}[b]
  \centering
    \includegraphics[width=\linewidth]{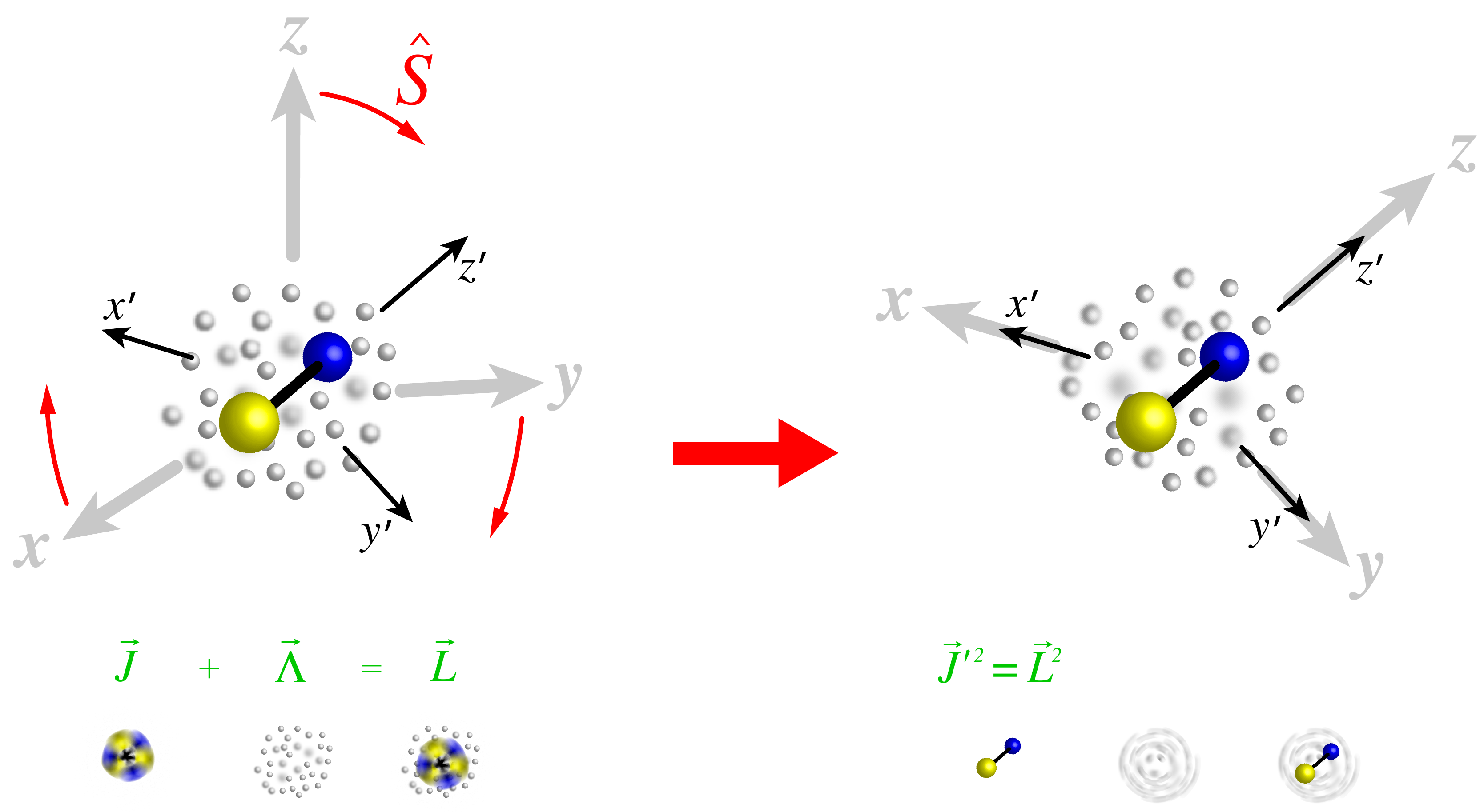}
  \caption{\label{transf} Action of the canonical transformation, Eq. (\ref{Transformation}), on the many-body system. Left: in the laboratory frame, $(x, y, z)$, the molecular angular momentum, $\mathbf{J}$, combines with the bath angular momentum, $\boldsymbol{\Lambda}$, to form the total angular momentum of the system, $\mathbf{L}$. Right: after the transformation, the bath degrees of freedom are transferred to the rotating frame of the molecule, $(x', y', z')$. As a result, the molecular angular momentum in the transformed space coincides with the total angular momentum of the system in the laboratory frame.}
 \end{figure}

The second term of Eq.~(\ref{Hamil}) represents the kinetic energy of the bosonic bath, where the corresponding creation and annihilation operators, $\bed_\mathbf{k}$ and $\be_\mathbf{k}$, are expressed  in the spherical basis, $\bed_{k\lambda \mu}$ and $\be_{k\lambda \mu}$. Here $k=|\mathbf{k}|$, while $\lambda$ and $\mu$ define, respectively, the boson angular momentum and its projection onto the laboratory $z$-axis, see Appendix~\ref{sec:Bklm} for details.

The last term of Eq.~(\ref{Hamil}) describes the interaction between the impurity and the bath. The angular-momentum-dependent coupling strength, $U_\lambda(k)$, depends on the microscopic details of the two-body interaction between the impurity and the bosons. For example, in Ref.~\cite{SchmidtLem15} we showed that for a linear rotor immersed into a Bose gas, the couplings are given by
\begin{equation}
\label{Ulamk}
	U_\lambda(k) = u_\lambda \left[\frac{8 k^2\epsilon_k \rho }{\omega_k(2\lambda+1)}\right]^{1/2} \int dr r^2 f_\lambda(r) j_\lambda (kr).
\end{equation}
This assumes that in the impurity frame, the   interaction between the rotor and  a bosonic atom is expanded as
\begin{equation}
\label{VimpBos}
	V_\text{imp-bos} (\mathbf{r'}) = \sum_\lambda u_\lambda f_\lambda (r') Y_{\lambda 0} ( \Theta', \Phi' ),
\end{equation}
with $u_\lambda$ and $f_\lambda(r')$ giving the strength and shape of the potential in the corresponding  angular momentum channel. The prefactor of Eq. (\ref{Ulamk}) depends on the bath density, $\rho$, the kinetic energy of the bare atoms, $\epsilon_k$, and the dispersion relation  of the bosonic quasiparticles, $\omega_k$. 
Since the angulon Hamiltonian (\ref{Hamil}) describes the interactions between a quantum rotor and a  bosonic bath of, in principle, any kind, we will approach it from an entirely general perspective, exemplifying the couplings by the ones of Eq. (\ref{Ulamk}).

Many-body problems such as given by the Hamiltonian~\eqref{Hamil} are typically hard to solve. The conventional approaches to tackle them include, when applicable, perturbation theory, renormalization group, or in principle uncontrolled methods such as  those based on the selective diagram resummations, as well as purely numerical techniques. An alternative, actively used since the development  of classical mechanics, involves canonical transformations of the underlying Hamiltonian~\cite{WagnerUnitaryBook, LLI}. Here the idea is to partially diagonalize  the Hamiltonian and/or to  expose the constants of motion, which allows to reveal some of the eigenstates' properties exactly. In the context of impurity problems, typical approaches  employ the collective bath variables as a generator of the symmetry transformations, as it has been used e.g. in the polaron theory \cite{Lee1953, GirardeauPF61, Devreese13}. 

In the angulon problem discussed in this paper, the total angular momentum is a good quantum number. However, due to the coupling of bath degrees of freedom with the impurity coordinates, as given by the third term of Eq.~\eqref{Hamil},  this conservation law is not apparent. Here we introduce a canonical transformation which   makes this constant of motion explicit and allows to achieve several other goals listed below. The corresponding operator $\hat S$ uses  the composite angular momentum of the bath as a generator of rotation, which transfers the environment degrees of freedom into the frame co-rotating along with the quantum rotor. The transformation is given by:
\begin{equation}
\label{Transformation}
	\fbox{$ \hat{S} = e^{- i \hat\phi \otimes \hat \Lambda_z} e^{- i \hat\theta  \otimes \hat\Lambda_y} e^{- i \hat\gamma  \otimes\hat \Lambda_z} \\$}
\end{equation}
The angle operators,  $(\hat\phi, \hat\theta, \hat\gamma)$, act  in the  Hilbert space of the rotor, and
\begin{equation}
\label{Lambda}
	 \hat {\vec\Lambda}=\sum_{k\lambda\mu\nu}\bed_{k\lambda\mu}\boldsymbol\sigma^{\lambda}_{\mu\nu}\be_{k\lambda \nu}
\end{equation}
is the collective angular momentum operator of the many-body bath, acting in the  Hilbert space of the bosons. Here $\boldsymbol\sigma^{\lambda}$ denotes the vector of matrices fulfilling the angular momentum algebra in the representation of angular momentum $\lambda$.

The transformation brings the Hamiltonian (\ref{Hamil}) into the following form:
\begin{multline}
\label{transH}
\hat{\mathcal{H}} \equiv \hat S^{-1} \hat H \hat S= B (\hat{\mathbf{J}}' - \hat{\mathbf{\Lambda}})^2 \\ + \sum_{k\lambda\mu}\omega_k \bed_{k\lambda\mu}\be_{k\lambda\mu} + \sum_{k\lambda} V_\lambda(k) \left[\bed_{k\lambda0}+\be_{k\lambda0}\right]
\end{multline}
Here $V_\lambda(k)=U_\lambda(k) \sqrt{(2\lambda+1)/(4\pi)}$ and $\hat{\mathbf{J}}'$ is the `anomalous' angular momentum operator acting in the rotating frame of the impurity. Since the components of $\hat{\mathbf{J}}'$ act in the body-fixed frame, they obey anomalous commutation relations~\cite{BiedenharnAngMom, LevebvreBrionField2} as opposed to the `ordinary' angular momentum operator, $\hat{\mathbf{J}}$ of Eq.~(\ref{Hamil}), which acts in the laboratory frame. The details of the derivation, as well as the properties of the $\hat{\mathbf{J}}'$ operator are presented in  Appendix~\ref{app:transfo}.

 Let us now discuss the physical meaning of the transformation $\hat S$. In order to describe the composite system, it is natural to introduce two coordinate frames, as schematically shown in Fig.~\ref{transf}. The laboratory frame, $(x,y,z)$, is singled out by the collective state of the bosons, while the rotating impurity frame, $(x',y',z')$, is defined by the instantaneous orientation of the molecular axes. The relative orientation of the two frames is given by the eigenvalues of the Euler angle operators, $(\hat\phi, \hat\theta, \hat\gamma)$, acting in the impurity Hilbert space. The $\hat S$ operator transforms the many-body state of the bosons into the rotating molecular frame, using $\hat{\vec \Lambda}$ as a generator of quantum rotations. In turn, as we show below, the molecular state  in the transformed frame becomes an eigenstate of the total angular momentum of the system, which is a constant of motion.

Introducing the body-fixed coordinate frame bound to the impurity makes explicit an additional quantum number, $n$, which gives the projection of the angular momentum onto the rotor axis $z'$. The angular momentum basis states, $\vert j, m, n \rangle$, are therefore the eigenstates of the $\hat{\mathbf{J}}^2$, $\hat{J}_z$, and $\hat{J}'_z$ operators, as given by Eqs.~(\ref{J2eig})--(\ref{JPrzeig}) of the Appendix.

For a linear-rotor molecule in the absence of a bath the total angular momentum, $\hat{\mathbf{L}} = \hat{\mathbf{J}} + \hat{\mathbf{\Lambda}}$, coincides with $\hat{\mathbf{J}}$. Therefore, $\hat{\mathbf{L}}$ is perpendicular to the molecular axis $z'$, resulting in $n=0$. With the bosons present, the total angular momentum is no longer perpendicular to $z'$, providing the molecular state with nonzero $n$ in the transformed frame. In other words, the transformation (\ref{Transformation}) converts a linear-rotor molecule into an effective `symmetric top'~\cite{LevebvreBrionField2} by dressing it with a boson field.

Compared to the original Hamiltonian, Eq.~(\ref{Hamil}), the transformed Hamiltonian, Eq.~(\ref{transH}), possesses the following properties:
\begin{enumerate}
\item
 \textit{$\hat{\mathcal{H}}$ is explicitly expressed through the total angular momentum, which is a constant of motion.} Due to the isotropy of space, the eigenstates of the original Hamiltonian, $\hat H$, are simultaneous eigenstates of the total angular momentum operators, $\hat{\mathbf{L}}^2$ and $\hat{\mathbf{L}}_z$, and thus can be labeled as $\vert L, M \rangle$. The transformed states, $\hat S^{-1} \vert L, M \rangle$, are hence the  eigenstates of the transformed Hamiltonian, $\hat{\mathcal{H}}$. 
 As detailed in Appendix~\ref{app:moltrans}, these transformed states are also eigenstates of the $\hat{\mathbf{J}}'^2$ operator with the eigenvalues $L(L+1)$, corresponding to the total angular momentum. Consequently, the $\hat{\mathbf{J}}'^2$ operator  in Eq. (\ref{transH}) can be replaced by the classical number $L(L+1)$.

 \item
 \label{2lab}
\textit{$\hat{\mathcal{H}}$ does not contain the impurity coordinates $(\hat \theta,\hat \phi)$, which allows to bypass the intractable angular momentum algebra, arising from the impurity-bath coupling.} The angle operators of the original Hamiltonian, Eq.~(\ref{Hamil}), couple the impurity states with every single boson excitation, which results in the problem of adding an infinite number of angular momenta in three dimensions. The latter involves working with Wigner $3nj$-symbols of an arbitrarily large order. In the transformed Hamiltonian, on the other hand, the problem is reduced to adding the angular-momentum projections of the impurity and the bath. There, the impurity-bath coupling, $\hat{\mathbf{J}}' \cdot \hat{\mathbf{\Lambda}}$, has the form of spin-orbit interaction and does not lead to an involved angular momentum algebra.

\item
\textit{$\hat{\mathcal{H}}$ can be solved exactly in the limit of a slowly rotating impurity, $B\to 0$}, see Sec.~\ref{sec:deform}.

\item
\textit{$\hat{\mathcal{H}}$ allows to find the eigenstates containing an infinite number of phonon excitations, which is crucial e.g.\ to account for the macroscopic deformation of the condensate.}  This follows directly from sub.~\ref{2lab}, and is detailed in Sec.~\ref{sec:deform}.

\item
\textit{$\hat{\mathcal{H}}$ contains information about the  deformation of the condensate in the rotating impurity frame.}  Compared to the laboratory frame, where  the deformation of the bath is averaged over the angles, this provides an additional insight into the nature of the many-body state and, consequently, into the origin of the angulon instability, discussed in Sec.~\ref{sec:deform}.

\end{enumerate}

  \begin{figure}[t]
  \centering
    \includegraphics[width=\linewidth]{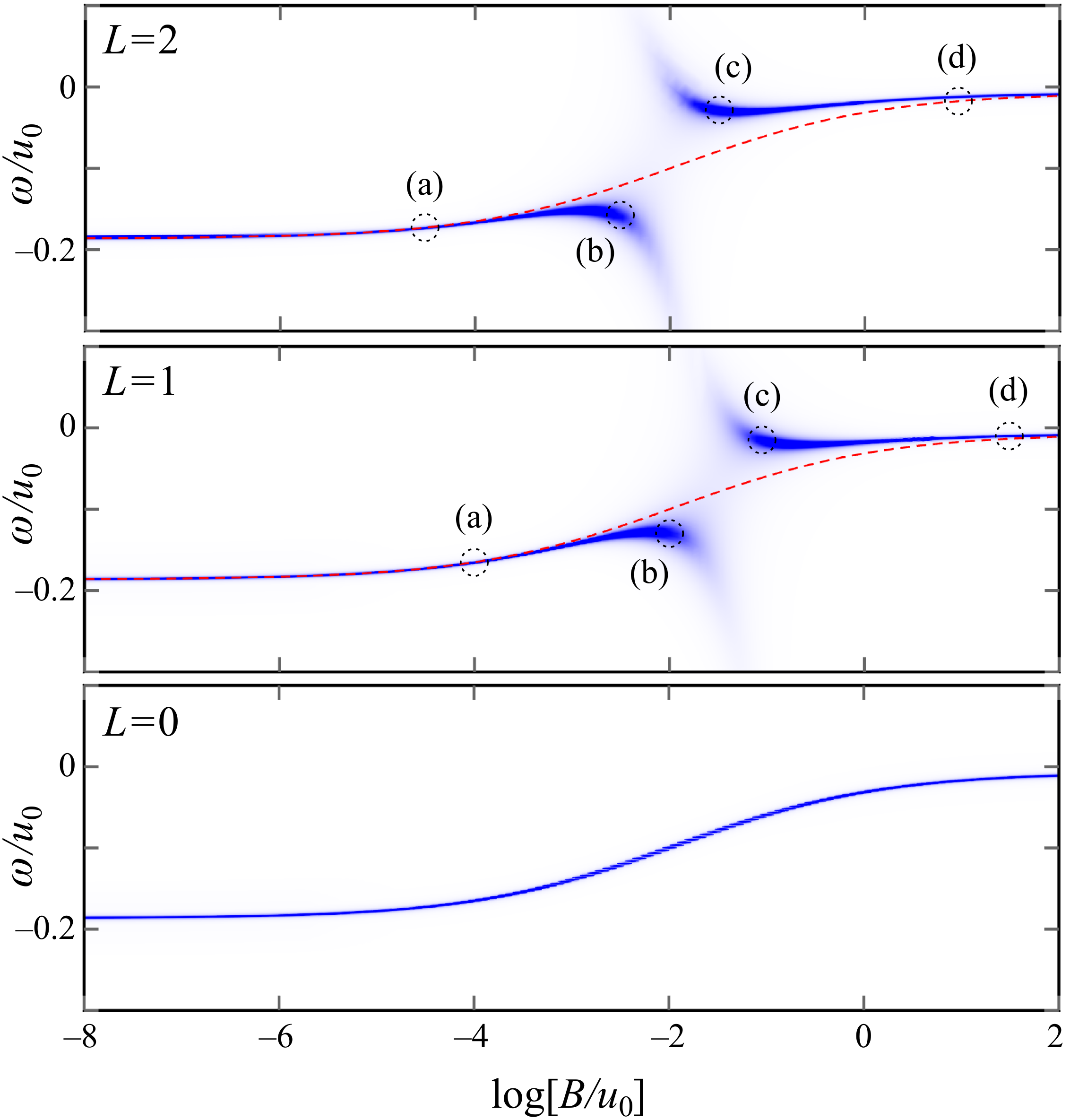}
  \caption{\label{energies} Change of the angulon spectral function, $A_L (\omega)$, where $\omega=E-B L(L+1)$, with the rotational constant $B$, for three lowest total angular momentum states. The $L>0$ states show an instability in the spectrum. The red dashed line shows the deformation energy, Eq.~(\ref{defEnergy}), which is independent of $L$. The circles indicate the points for which the phonon density modulation is shown in Fig.~\ref{PhonDens}.}
 \end{figure}

\section{Macroscopic deformation of the bath and the emerging instability}
\label{sec:deform}

In the limit of a slowly-rotating impurity, $B \to 0$, the Hamiltonian (\ref{transH}) can be solved exactly by means of  an additional canonical transformation:
\begin{equation}
\label{Htilde}
 \hat{\mathscr{H}} = \hat{U}^{-1} \hat{\mathcal{H}} \hat{U}
\end{equation}
where
\begin{equation}
\label{Utransf}
	\hat{U} = \exp \left[ \sum_{k \lambda} \frac{V_\lambda (k)}{W_{k\lambda}} \left( \be_{k \lambda 0} - \bed_{k \lambda 0}  \right) \right]
\end{equation}
with $W_{k\lambda}=\omega_k+B\lambda(\lambda+1)$. This transformation removes the terms linear in the bosonic operators, replacing them by the deformation energy of the bath,
 \begin{equation}
\label{defEnergy}
	E_\text{def} = - \sum_{k\lambda} V_\lambda(k)^2/W_{k\lambda}
\end{equation}
As a consequence, in the limit of $B=0$, the vacuum of phonon excitations, $\ket{0}$, becomes the exact ground state of Eq.~(\ref{Htilde}). On the other hand, such a coherent shift transformation corresponds to a macroscopic deformation of the bath, and could not be easily performed on the original Hamiltonian (\ref{Hamil})  where the impurity coordinates are strongly coupled with the bath degrees of freedom. 

Here we are interested in the effect of a slowly-rotating impurity on the many-body state of the environment. Therefore we introduce a variational ansatz based on single-phonon excitations on top of the bosonic state macroscopically deformed by the operator $\hat U$:
\begin{equation}
\label{PsiChevy}
	\vert \psi \rangle = g_{LM} \vert 0 \rangle \vert L M 0 \rangle + \sum_{k \lambda n}  \alpha_{k \lambda n}   \bed_{k \lambda n} \vert 0 \rangle \vert L M n \rangle
\end{equation}
The states of an isolated symmetric-top molecule are characterised by three quantum numbers: the angular momentum, $L$, its projection, $M$, onto the laboratory-frame $z$-axis, and its projection, $n$, onto the molecular symmetry axis, $z'$. For a linear rotor molecule, the angular momentum vector is always perpendicular to the molecular axis and therefore $n$ is identically zero. The transformation~(\ref{Transformation}), however, transfers the bosons to the molecular frame, thereby   creating an effective `many-body symmetric-top' state. The latter consists of a linear rotor impurity dressed by the field of bosons carrying finite angular momentum. As a result, the total angular momentum of such a symmetric-top is no longer perpendicular to the linear rotor axis and provides the finite values of the projection $n$. See Appendix~\ref{app:moltrans} for more details.

It is worth emphasizing that the non-transformed  many-body wavefunction corresponding to Eq.~(\ref{PsiChevy}) is given by $\ket{\phi}= \hat S \cdot \hat U\ket{\psi}$. Therefore it is a highly-involved object with an infinite number of degrees of freedom entangled with each other. The simple ansatz of Eq.~(\ref{PsiChevy}) was made possible by the consecutive canonical transformations, Eqs.~(\ref{Transformation}) and (\ref{Utransf}). Furthermore, it is straightforward to extend Eq.~(\ref{PsiChevy}) to bath excitations of higher order, since this does not generate any complexities related to the angular momentum algebra.

Performing the variational solution for the energy, $E = \langle \psi \vert  \hat{\mathscr{H}} \vert \psi \rangle/ \langle \psi \vert  \psi \rangle$, we obtain the condition
\begin{equation}
\label{Dyson}
	-E+BL(L+1)-\Sigma_L(E)=0
\end{equation}
which has the form of a Dyson equation with self-energy $\Sigma_L(E)$ \cite{Rath2013}, as given by Eq.~(\ref{selfenergyapp}), see Appendix~\ref{sec:variational} for a detailed derivation. Eq.~(\ref{Dyson}) can be rewritten in terms of the angulon Green's function as $\left[G_L(E)\right]^{-1}=0$, where
\begin{equation}
\label{GFct}
\left[G_L(E)\right]^{-1}= [G^0_L(E)]^{-1} - \Sigma_L(E) ,
\end{equation}
with  $[G^0_L(E)]^{-1} =-E+B L(L+1)$.

   \begin{figure}[hbtp]
  \centering
    \includegraphics[width=\linewidth]{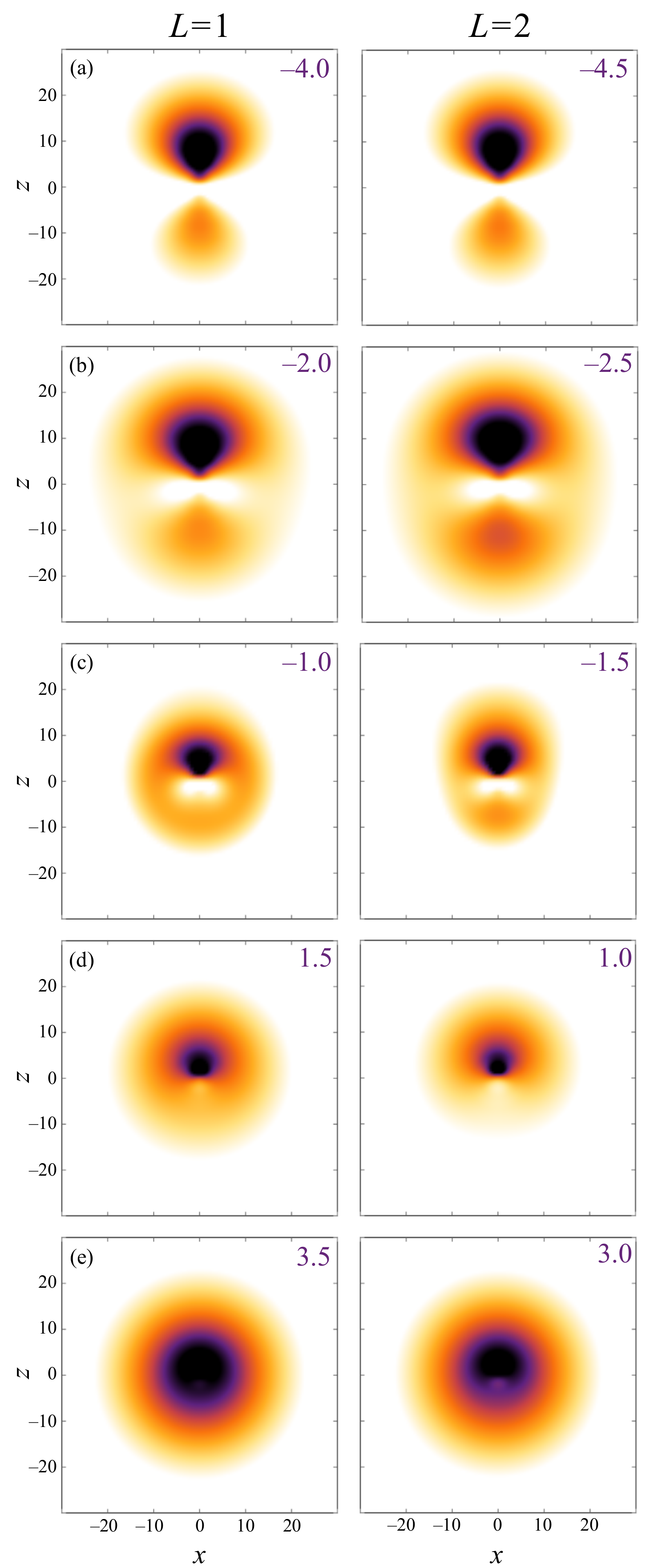}
  \caption{\label{PhonDens} Phonon density in the impurity frame for selected values of $\log [B/u_0]$, which are specified in the right-top corner of the panels and (a)--(d) as labeled in Fig.~\ref{energies}; (e) far to the right from the instability, at $\log[B/u_0] = 3.5$ for $L=1$ and at $\log[B/u_0]  = 3.0$ for $L=2$. The coordinates $(x,z)$ are in units of $(m u_0)^{-1/2}$.}
 \end{figure}

The ground- and excited-state properties of the system are contained in the spectral function, $\mathcal A_L (E)= \text{Im}[G_L(E + i 0^+)]$. Without restricting the generality of what follows, we assume potentials whose angular momentum expansion, Eq.~(\ref{VimpBos}), is given by the Gaussian form-factors,  $f_\lambda(r)=(2\pi)^{-3/2}e^{-r^2/(2r_\lambda^2)}$, and nonzero magnitudes, $u_0$ and $u_1$, in two lowest angular momentum channels. We assume an anisotropy ratio of $u_1/u_0=5$, a range $r_0 = r_1 = 15~(m u_0)^{-1/2}$, and set the interactions with $\lambda>1$ to zero. Furthermore, we use a Bogoliubov-type dispersion relation, $\omega_k  = \sqrt{\epsilon_k (\epsilon_k + 2 g_\text{bb} n)}$, where $\epsilon_k = k^2/2 m$ with $m$ the mass of a boson. We choose the boson-boson interaction $g_\text{bb} = 418 (m^3 u_0)^{-1/2}$ and  density $n = 0.014 (m u_0)^{3/2}$. This choice of parameters reproduces the speed of sound in superfluid $^4$He for $u_0 = 2\pi \times100$~GHz \cite{DonnellyHe98}. Fig.~\ref{energies} shows the dependence of the spectral function on the rotational constant $B$ for the three lowest rotational states.  The width of the lines reflects the lifetimes of the corresponding levels.  In Ref.~\cite{SchmidtLem15}, we studied the non-transformed Hamiltonian~\eqref{Hamil} using a variational ansatz based on single-phonon excitations. Using this ansatz we found that the angulon states become stable after crossing the phonon threshold at zero energy. Here this is no longer the case, since the transformation $\hat U$ of eq.~\eqref{Utransf} introduces an infinite number of phonon excitations into the variational ansatz. This leads to a energetic renormalization of the phonon emission threshold providing all the excited angulon states with decay channels for phonon emission. This, in turn, leads to a finite lifetime for any magnitude of the impurity-bath coupling.

In the limit of $B \to 0$ the molecule is not rotating and is inducing an anisotropic deformation of the bath, corresponding to the mean-field-like deformation energy, Eq.~(\ref{defEnergy}). The magnitude of the deformation energy decreases with $B$ monotonously and determines the general shape of the spectrum. Apart from the  deformation energy which is identical for all $L$'s, the energy of the angulon acquires an additional contribution due to phonon excitations in the surrounding medium. The latter corresponds to the rotational Lamb shift discussed in Ref.~\cite{SchmidtLem15}, which has been observed as the renormalization of the rotational spectrum for molecules in superfluid helium nanodroplets~\cite{ToenniesAngChem04}. Most importantly, we find that for the excited states with $L>0$ there exists a critical rotational constant, where a discontinuity in the rotational spectrum occurs. This effect corresponds to a transfer of one quantum of angular momentum \textit{from the bath to the impurity}. One can see that the faster the rotation (i.e.\ the larger $L$), the earlier this instability occurs.  
Such an instability has been briefly discussed in Ref.~\cite{SchmidtLem15}, where it was referred to as Many-Body-Induced Fine Structure of the second kind.

While the instability can be detected using spectroscopy in the laboratory frame, an insight into its origin can be gained by making use of the canonical transformation, Eq.~(\ref{Transformation}). Namely, in the frame co-rotating with the impurity, the instability manifests itself as a change of the phonon density, $\langle \bd_\mathbf{r} \be_\mathbf{r} \rangle$; for analytic expressions see Appendix~\ref{sec:deformation}. Fig.~\ref{PhonDens} shows the phonon density  for $L=1$ and $2$ at five different values of the impurity rotational constant. Darker shade corresponds to higher density. Far to the left of the instability, panels~(a), the impurity is rotating slowly and the bosons are able to adiabatically follow its motion. As a result, the surrounding bath becomes  polarized, which manifests itself in a highly-asymmetric phonon density. The shape of the density modulation is given by the first spherical harmonic which arises due to the $\lambda=1$ term in the impurity-boson potential included in our model. Closer to the instability, panels~(b), the phonon density increases, signaling the onset of the resonant phonon excitations. At the right edge of the instability, panels~(c), the phonon density drops drastically. Further away from the instability, the density distribution becomes the more symmetric the faster the impurity rotates, as illustrated in panels (d), (e).  In other words, when the rotational constant exceeds the critical value given by the instability, it becomes energetically unfavorable for the bosons to follow the motion of the impurity. As a consequence, the bosonic bath does not possess finite angular momentum, which results in the spherically-symmetric density distribution. Thus, the phonon density in the transformed frame can serve as a fingerprint of the angular momentum transfer from the bath to the impurity which takes place at the instability point.

It is important to note that the `angulon instability' discussed here is fundamentally different from the vortex instability~\cite{Pitaevskii2003}, also associated with rotation. The comparison between the two is summarized in Table~\ref{tab:inst}. 
First, the rotation of the impurity is inherently three-dimensional and does not involve any specific rotation axis. This is different for a vortex line which singles out a particular direction in space.  Second, the formation of a vortex requires a transfer of one unit of angular momentum per particle in the bath. In the angulon instability, on the other hand, a finite (small) number of rotational quanta is shared between the impurity and the collective state of the many-particle environment. Finally, the vortex instability leads to a finite circulation around the vortex line, which is absent for the angulon instability.

\begin{table}[t]
\caption{\label{tab:inst} Comparison of the angulon instability with the vortex instability.}
\begin{tabular}{ | c | c | c |}
\hline
   & Angulon & Vortex \\
\hline
  Corresponding rotation & Spherical, $\hat{\mathbf{L}}^2$ & Planar, $\hat{\mathbf{L}}_z$  \\[2pt]
Angular momentum transfer & $\hbar$ & $\hbar$ per particle \\[2pt]
Circulation & zero & integer$\times 2 \pi \hbar/m$ \\[2pt]
\hline
\end{tabular}
\end{table}

 \section{Experimental implementation}

   \begin{figure}[b]
  \centering
    \includegraphics[width=\linewidth]{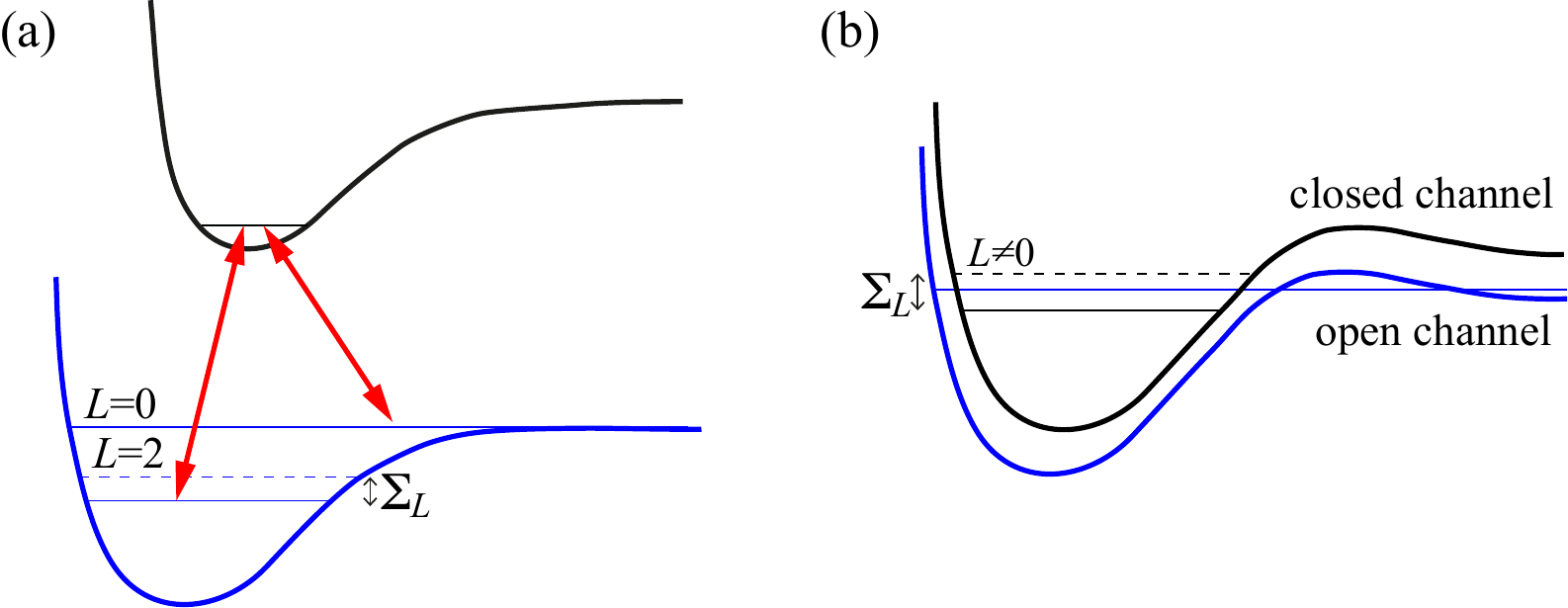}
  \caption{\label{PA} Detection of the angulon self-energy $\Sigma_L$ using (a)~photoassociation spectroscopy~\cite{UlmanisCR12} and (b)~shift of $p$- and $d$-wave Feshbach resonances~\cite{KohlerRMP06}.}
 \end{figure}
 
The described effects can be observed  experimentally both with molecules trapped in strongly-interacting superfluids, such as helium droplets~\cite{ToenniesAngChem04}, or molecular impurities immersed in weakly-interacting Bose-Einstein condensates~\cite{Pitaevskii2003}. The dependence of the angulon self-energy, $\Sigma_L$ of Eq.~\eqref{GFct}, on the many-body parameters can be revealed by measuring the relative shift between the rotational states of a diatomic molecule. Since the effects will be most pronounced for the molecular states possessing  a small rotational constant $B$, experiments involving molecules in highly-excited vibrational states provide the most natural setup. In the context of ultracold gases, the latter  include  photoassociation spectroscopy~\cite{UlmanisCR12} and measuring nonzero angular momentum Feshbach resonances~\cite{KohlerRMP06}. In both cases, the shifts of the spectroscopic lines  will be proportional to the angulon self-energy, as schematically illustrated in Fig.~\ref{PA}. An alternative possibility is measuring $\Sigma_L$ as a shift of the microwave lines in the spectra of weakly bound molecules~\cite{MarkPRA07}, prepared using one of these techniques. In frequency domain, at sufficiently low temperatures the width of the lines will correspond to the angulon lifetime. The instability shown in Fig.~\ref{energies} corresponds to the vanishing quasiparticle weight with a related emergence of a broad incoherent background and therefore can be detected as a  line broadening  with increasing impurity-bath interactions.
 In the time domain, on the other hand, the angulon Green's function can be detected using  Ramsey and spin-echo techniques~\cite{KnapPRX12, Schmidt15}. In such measurement, the angulon instability will leads to dephasing dynamics with a related pronounced decay of the Ramsey and spin-echo contrast~\cite{KnapPRX12, Schmidt15}.

While in superfluid helium the interactions cannot be tuned as easily as in ultracold gases, the range of chemical species amenable to trapping is essentially unlimited~\cite{ToenniesAngChem04}. The latter, combined with the advances in the theory of molecule-helium interactions~\cite{SzalewiczIRPC08} paves the way to studying angulon physics in a broad range of parameters.

\section{Conclusions}

In this paper, we studied the redistribution of orbital angular momentum between a quantum impurity and a many-particle environment. We introduced a technique which allows to drastically simplify the problem of adding an infinite number of angular momenta which occurs in the regime of strong interactions. The essence of the method -- a novel canonical transformation -- paves the way to eliminating the complex angular momentum algebra from the problem, as well as to exposing the problem's constants of motion.  We exemplified the technique's capacity by studying an instability which occurs in the spectrum of the many-particle system due the interaction between the bath and the rotating impurity. Such an instability should be detectable with molecules in superfluid helium droplets~\cite{ToenniesAngChem04} and might be responsible for the long timescales emerging in molecular rotation dynamics in the presence of an environment~\cite{PentlehnerPRL13}, which presently lacks even a qualitative explanation. Moreover, the rotating impurities can be prepared experimentally in perfectly controllable settings, based on ultracold molecules immersed into a Bose or Fermi gas~\cite{JinYeCRev12, KreStwFrieColdMol, LemKreDoyKais13} and cold molecular ions inside Coulomb crystals~\cite{WillitschIRPC12}.  It is important to note that the transformation, as defined by Eq.~\eqref{Transformation}, is quite general, and can be applied to extended Fr\"ohlich Hamiltonians \cite{Rath2013}, to impurities with complex rotational structure~\cite{LevebvreBrionField2}, Rydberg molecules \cite{Greene2000,bendkowsky2009, BellosPRL13, KruppPRL14}, as well as to the case of a Fermionic bath~\cite{ChevyPRA06}.

The ultimate goal of our approach is to find a series of canonical transformations that would lead to exact solutions to the many-body Hamiltonians of the same class as Eq.~(\ref{Hamil}). This resonates with Wegner's idea of the continuous unitary transformations~\cite{WegnerAnnPhys94}, which underlies one of the Hamiltonian formulations of the renormalization group approach~\cite{KehreinRG}.

Finally, the impurity problem considered here can be used as a building block of a general theory describing the redistribution of orbital angular momentum in quantum many-particle systems. This opens up a perspective of applying the techniques of this article to the several problems in condensed matter~\cite{Mahan90} and chemical~\cite{EncyclChemPhys} physics.

\vspace{0.5cm}

\section{Acknowledgements} 

We are grateful to Eugene Demler, Jan Kaczmarczyk, Laleh Safari, and Hendrik Weimer for insightful discussions. The work was supported by the NSF through a grant for the Institute for Theoretical Atomic, Molecular, and Optical Physics at Harvard University and Smithsonian Astrophysical Observatory.

\appendix

\section{The angular momentum representation}
\label{sec:Bklm}

The creation and annihilation operators of Eq.~(\ref{Hamil}) are expressed in the angular momentum representation, which is related to the Cartesian representation as:
\begin{equation}
\label{AklmAk}
	\bd_{k\lambda \mu} =\frac{k}{(2\pi)^{3/2}} \int   d\Phi_k d\Theta_k~\sin\Theta_k~\bd_\mathbf{k} ~i^\lambda~ Y^*_{\lambda \mu} (\Theta_k, \Phi_k)
\end{equation}
\begin{equation}
\label{AkAklm}
	\bd_\mathbf{k} = \frac{(2\pi)^{3/2}}{k} \sum_{\lambda \mu}  \bd_{k\lambda \mu}~i^{-\lambda}~Y_{\lambda \mu} (\Theta_k, \Phi_k), 
\end{equation}
The quantum numbers $\lambda$ and $\mu$ define, respectively, the angular momentum of the bosonic excitation and its projection onto the laboratory-frame $z$-axis. Eqs.~(\ref{AklmAk}) and~(\ref{AkAklm})  correspond to the following commutation relations:
\begin{equation}
\label{AkComm}
	[\be_\mathbf{k}, \bd_\mathbf{k'}] = (2\pi)^3\delta^{(3)}(\mathbf{k-k'})
\end{equation}
\begin{equation}
\label{AklmComm}
	[\be_{k\lambda \mu}, \bd_{k'\lambda' \mu'}] = \delta(k-k') \delta_{\lambda \lambda'} \delta_{\mu \mu'}
\end{equation}

In the coordinate space, the transformation between the representations is defined as:
\begin{equation}
\label{ArlmAr}
	\bd_{r\lambda \mu} = r  \int d\Phi_r d\Theta_r~\sin\Theta_r ~\bd_\mathbf{r}~i^\lambda~Y^*_{\lambda \mu} (\Theta_r,\Phi_r)
\end{equation}
\begin{equation}
\label{ArArlm}
	\bd_\mathbf{r} = \frac{1}{r} \sum_{\lambda \mu}  \bd_{r\lambda \mu} ~i^{-\lambda}~Y_{\lambda \mu} (\Theta_r,\Phi_r)
\end{equation}
with the corresponding commutation relations:
\begin{equation}
\label{ArComm}
	[\be_\mathbf{r}, \bd_\mathbf{r'}] = \delta^{(3)}(\mathbf{r-r'})
\end{equation}
\begin{equation}
\label{ArlmComm}
	[\be_{r \lambda \mu}, \bd_{r' \lambda' \mu'}] = \delta(r-r') \delta_{\lambda  \lambda'} \delta_{\mu \mu'}
\end{equation}

The operators in the coordinate and momentum space are related through the Fourier transform,
\begin{equation}
\label{brViabk}
	\be^\dagger_\mathbf{r} = \int \frac{d^3  k}{(2\pi)^3} \be^\dagger_\mathbf{k} e^{i \mathbf{k \cdot r}},
\end{equation}
from which one can obtain the corresponding relation for the angular momentum components
\begin{equation}
\label{brlmViabklm}
	\be^\dagger_{r \lambda \mu} = i^\lambda \sqrt{\frac{2}{\pi}} r \int k dk~j_\lambda (kr)~\be^\dagger_{k\lambda\mu}
\end{equation}
with $j_\lambda (kr)$ the spherical Bessel function~\cite{AbramowitzStegun}.

\section{The canonical transformation}
\label{app:transfo}

Here we provide details on the derivation of the transformed Hamiltonian, Eq. (\ref{transH}).

In the angular momentum representation, the boson creation and annihilation operators, $\bed_{k\lambda \mu}$ and  $\be_{k\lambda \mu}$, are defined as irreducible tensors of rank $\lambda$~\cite{VarshalovichAngMom}. Consequently, they are transformed by  the $\hat{S}$-operator of Eq. (\ref{Transformation}) in the following way:
\begin{align}
\label{bDagRot}
	 \hat{S}^{-1}  \bed_{k\lambda \mu}  \hat{S} &= \sum_\nu  D_{\mu \nu}^{\lambda \ast} (\hat{\phi}, \hat{\theta}, \hat{\gamma})   \bed_{k\lambda \nu}\\
\label{bRot}
	 \hat{S}^{-1}  \be_{k\lambda \mu}  \hat{S} &= \sum_\nu  D_{\mu \nu}^{\lambda} ( \hat{\phi},  \hat{\theta},  \hat{\gamma})  \be_{k\lambda \nu}
\end{align}
Here $D_{\mu \nu}^{\lambda} ( \hat{\phi},  \hat{\theta},  \hat{\gamma})$ are Wigner $D$-matrices~\cite{VarshalovichAngMom} whose arguments are the angle operators defining the relative orientation of the impurity frame with respect to the laboratory frame.
These expressions can also be derived using the explicit expression for the angular momentum of the bosons, Eq.~(\ref{Lambda}).

The Wigner rotation matrix appearing in Eq. (\ref{bDagRot}) is complex conjugate with respect to the one of Eq. (\ref{bRot}) and therefore corresponds to an inverse rotation. As a result, 
\begin{equation}
\label{bDagbRot}
 	\hat{S}^{-1} \Bigl( \sum_\mu  \bed_{k\lambda \mu}  \be_{k\lambda \mu}  \Bigr) \hat{S} =    \sum_\mu \bed_{k\lambda \mu}  \be_{k\lambda \mu} ,
\end{equation}
and the second term of Eq.~(\ref{Hamil}) does not change under the transformation.

Similarly, in the last term of Eq.~(\ref{Hamil}) we use that $Y_{\lambda \mu} (\hat{\theta}, \hat{\phi}) = \sqrt{\frac{2 \lambda +1}{ 4 \pi}}  D^{\lambda \ast}_{\mu 0} (\hat{\phi}, \hat{\theta}, 0)$, which leads to cancellation of the Wigner $D$-matrices. In such a way, the transformation $\hat{S}$ eliminates the molecular angle variables from the Hamiltonian.

The transformation of the molecular rotational Hamiltonian, $B \mathbf{\hat{J}^2}$, turns out to be slightly more cumbersome. In the laboratory frame, the angular momentum vector is defined by its spherical components, $\hat{\mathbf{J}} = \{\hat{J}_{-1}, \hat{J}_{0}, \hat{J}_{+1}  \}$, where:
\begin{align}
\label{J0}
	 \hat{J}_0 &= \hat{J}_z \\
 \label{Jplus}
	 \hat{J}_{+1} &= -\frac{1}{\sqrt{2}} \left(\hat{J}_x + i\hat{J}_y \right)\\
\label{Jminus}
	 \hat{J}_{-1} &= \frac{1}{\sqrt{2}} \left(\hat{J}_x - i\hat{J}_y \right)
\end{align}
see Refs.~\cite{BiedenharnAngMom, VarshalovichAngMom}. We use the analogous notation  for the components of the total angular momentum of the bosons $\hat{\boldsymbol{\Lambda}} = \{\hat{\Lambda}_{-1}, \hat{\Lambda}_{0}, \hat{\Lambda}_{+1}  \}$, Eq.\ (\ref{Lambda}). The operators (\ref{J0})--(\ref{Jminus}) obey the following commutation relations with each other:
\begin{equation}
\label{Jicommute}
 	\left[\hat{J}_i, \hat{J}_k \right] = - \sqrt{2} C_{1, i; 1, k}^{1, i+k}  \hat{J}_{i+k},
\end{equation}
where $i,k = \{-1, 0, +1 \}$, and with the rotation operators:
\begin{multline}
\label{JiComm}
 	\left[\hat{J}_k, D^{\lambda}_{\mu \nu}  (\hat{\phi}, \hat{\theta}, \hat{\gamma}) \right]  \\= (-1)^{k+1} \sqrt{\lambda(\lambda+1)} C_{\lambda, \mu; 1,-k}^{\lambda, \mu-k} D^{\lambda}_{\mu-k, \nu}  (\hat{\phi}, \hat{\theta},  \hat{\gamma})
\end{multline}
\begin{multline}
\label{JiCommAst}
 	\left[\hat{J}_k, D^{\lambda \ast}_{\mu \nu}  (\hat{\phi}, \hat{\theta}, \hat{\gamma}) \right]  \\=  \sqrt{\lambda(\lambda+1)} C_{\lambda, \mu; 1, k}^{\lambda, \mu+k} D^{\lambda \ast}_{\mu+k, \nu}  (\hat{\phi}, \hat{\theta},  \hat{\gamma})
\end{multline}
Here $C_{l_1, m_1; l_2, m_2}^{l_3, m_3}$ are the Clebsch-Gordan coefficients~\cite{VarshalovichAngMom}. 

By using the latter property, one can show that the operators (\ref{J0})--(\ref{Jminus}) transform under Eq. (\ref{Transformation}) in the following way:
\begin{equation}
\label{JiTransformed}
 	\hat{\mathcal{J}}_{i} \equiv \hat{S}^{-1}  \hat{J}_{i} \hat{S} = \hat{J}_{i} - \sum_{k=-1,0,1} D^{1 \ast}_{i k} (\hat{\phi}, \hat{\theta}, \hat{\gamma}) \hat{\Lambda}_{k}
\end{equation}

After some angular momentum algebra, we obtain the following expression for the square of the angular momentum in the transformed frame:
\begin{equation}
\label{J2transformed}
	\hat{S}^{-1}  \mathbf{\hat{J}^2} \hat{S} \equiv \hat{\mathcal{J}}_{0}^2 - \hat{\mathcal{J}}_{+1} \hat{\mathcal{J}}_{-1} - \hat{\mathcal{J}}_{-1} \hat{\mathcal{J}}_{+1} =(\mathbf{\hat{J}'} - \mathbf{\hat{\Lambda}})^2 
\end{equation}

Here $\mathbf{\hat{J}'}$ is the angular momentum operator in the rotating molecular (i.e.~body-fixed) coordinate frame~\cite{LevebvreBrionField2, BiedenharnAngMom}, which can be expressed via the laboratory-frame components as:
\begin{equation}
\label{JiPrimeviaJi}
 	\hat{J}'_{i}   = \sum_k D^{1}_{k, i}  (\hat{\phi}, \hat{\theta}, \hat{\gamma}) \hat{J}_{k} 
\end{equation}
The spherical components of $\mathbf{\hat{J}'}$  are expressed through the Cartesian components using the relations analogous to Eqs.~(\ref{J0})-(\ref{Jminus}). Note that this makes the $\mathbf{\hat{J}'}$ operators different from the so-called contravariant angular momentum components used by Varshalovich~\cite{VarshalovichAngMom}.

The molecular-frame angular momentum operators obey the anomalous commutation relations with one another~\cite{BernathBook, BiedenharnAngMom},
\begin{equation}
\label{JiPrimecommute}
 	\left[\hat{J}'_i, \hat{J}'_k \right] = \sqrt{2} C_{1, i; 1, k}^{1, i+k}  \hat{J}'_{i+k}
\end{equation}
and the following commutation relations with the rotation matrices:
\begin{multline}
\label{JiPrimeComm}
 	\left[\hat{J}'_k, D^{\lambda}_{\mu \nu}  (\hat{\phi}, \hat{\theta}, \hat{\gamma}) \right]  \\= - \sqrt{\lambda(\lambda+1)} C_{\lambda, \nu; 1, k}^{\lambda, \nu+k} D^{\lambda}_{\mu, \nu + k}  (\hat{\phi}, \hat{\theta}, \hat{\gamma})
\end{multline}
\begin{multline}
\label{JiPrimeCommAst}
 	\left[\hat{J}'_k, D^{\lambda \ast}_{\mu \nu}  (\hat{\phi}, \hat{\theta}, \hat{\gamma}) \right]  \\= (-1)^k \sqrt{\lambda(\lambda+1)} C_{\lambda, \nu; 1, - k}^{\lambda, \nu-k} D^{\lambda}_{\mu, \nu - k}  (\hat{\phi}, \hat{\theta}, \hat{\gamma})
\end{multline}

It is worth noting that in the case of a linear-rotor molecule, the molecule-boson interaction does not depend on the third Euler angle, $\hat{\gamma}$. However, this angle must be preserved in Eq. (\ref{Transformation}), as well as in all the derivations described above, in order to keep the transformation unitary.

\section{Molecular states in the transformed space}
\label{app:moltrans}

In the main text and Fig.~\ref{transf} we have introduced two coordinate frames: the laboratory one, $(x,y,z)$, and the molecular one, $(x',y',z')$. A general molecular state, therefore, can be characterised by three quantum numbers: the magnitude of angular momentum, $j$; its projection, $m$, onto the laboratory-frame $z$-axis; and its projection, $n$, onto the molecular-frame $z'$-axis:
\begin{align}
\label{J2eig}
	\hat{\mathbf{J}}^2 \vert j, m, n \rangle &= j(j+1) \vert j, m, n \rangle \\
 \label{Jzeig}
	\hat{J}_z \vert j, m, n \rangle &= m \vert j, m, n \rangle \\
 \label{JPrzeig}
	\hat{J}'_z \vert j, m, n \rangle &= n \vert j, m, n \rangle
\end{align}

In the angular representation, the corresponding wave functions are given by~\cite{LevebvreBrionField2}:
\begin{equation}
\label{JMNwf}
 	\langle \phi, \theta, \gamma \vert j, m, n \rangle = \sqrt{\frac{2j+1}{8 \pi^2}} D^{j \ast}_{mn} (\phi, \theta, \gamma)
\end{equation}

The action of the space-fixed and molecule-fixed components of angular momentum is given by the general formula~\cite{BernathBook, BiedenharnAngMom}:
\begin{align}
\label{JiKet}
 	\hat{J}_k  \vert j, m, n \rangle &=  \sqrt{j(j+1)} C_{j, m; 1, k}^{j, m+k}  \vert j, m+k, n \rangle\\
\label{JiPrimeKet}
	\hat{J}'_k  \vert j, m, n \rangle &= (-1)^{k} \sqrt{j(j+1)} C_{j, n; 1, -k}^{j, n-k}  \vert j, m, n-k \rangle
\end{align}
where $k = \{-1, 0, +1\}$. Thus, in the molecular frame the raising operators lower the projection quantum number $n$ and the lowering operators raise it.

Unlike for nonlinear polyatomic molecules~\cite{LevebvreBrionField2}, the angular momentum of a linear rotor is always perpendicular to the internuclear axis (defining $z'$), and therefore $n$ is identically zero. However, this is the case only  before the transformation $\hat{S}$ is applied. Let us consider the most general many-body state in the non-transformed frame,
\begin{equation}
\label{LM}
 	 \vert L, M \rangle = \sum_{\underset{j m; i}{k \lambda \mu}} a_{k \lambda j}^i C_{j, m; \lambda, \mu}^{L, M} \vert j m 0 \rangle \otimes \vert k \lambda \mu \rangle_i
\end{equation}
The molecular states $\vert j m 0 \rangle$ are  the eigenstates of the molecular angular momentum operator, as given by Eqs. (\ref{J2eig})--(\ref{Jzeig}). The same relations are fulfilled for the collective bosonic states:  $\hat{\mathbf{\Lambda}}^2 \vert k \lambda \mu \rangle = \lambda(\lambda+1) \vert k \lambda \mu \rangle$ and $\hat{\mathbf{\Lambda}}_z \vert k \lambda \mu \rangle =\mu \vert k \lambda \mu \rangle$, where $\hat{\mathbf{\Lambda}}$ is defined by Eq. (\ref{Lambda}), and $k$ is the linear momentum. The index $i$ labels all the possible boson configurations resulting in a collective state $\vert k \lambda \mu \rangle$, spanning the complete many-body Hilbert space of the bosonic bath.

It is straightforward to show that the state (\ref{LM}) is an eigenstate of the total angular momentum operator, $\hat{\mathbf{L}} = \hat{\mathbf{J}}+\hat{\mathbf{\Lambda}}$:
\begin{align}
\label{L2eig}
	\hat{\mathbf{L}}^2\vert L, M \rangle &= L(L+1) \vert L, M \rangle \\
 \label{Lzeig}
	\hat{L}_z \vert L, M \rangle &= M \vert L, M \rangle
\end{align}

By acting on $ \vert L, M \rangle$ with $\hat{S}^{-1}$, after some angular momentum algebra, we obtain the state in the transformed frame:
\begin{equation}
\label{LMtil}
 	\hat{S}^{-1} \vert L, M \rangle = \sum_{k \lambda n i} f^i_{k \lambda n} \vert L M n \rangle \otimes \vert k \lambda n \rangle_i
\end{equation}
where the coefficients are given by
\begin{equation}
\label{fn}
 	f^i_{k \lambda n} = (-1)^{\lambda+n} \sum_{j}  a_{k \lambda j}^i C_{L, -n; \lambda, n}^{j, 0}
\end{equation}

We see that the transformation effectively transferred the angular momentum of the bosons to the molecular frame. This is reflected by the fact that the transformed state, $\hat{S}^{-1} \vert L, M \rangle$, becomes an eigenstate of the body-fixed angular momentum operator,  $\hat{\mathbf{J}}'^2$,  with the eigenvalues of the total angular momentum operator, $\hat{\mathbf{L}}^2$, i.e.
\begin{equation}
\label{J2Slm}
 	\hat{\mathbf{J}}'^2 \left( \hat{S}^{-1} \vert L, M \rangle \right)= L(L+1) \left( \hat{S}^{-1} \vert L, M \rangle \right)
\end{equation}

Each state $\vert L M n \rangle$ in the superposition of Eq.~(\ref{LMtil}) is an effective symmetric-top state~\cite{LevebvreBrionField2}, with the projection of total angular momentum on the molecular axis entirely determined by the boson field.

\section{Derivation of the Dyson equation from the variational principle}
\label{sec:variational}

We minimize the energy obtained from the expectation value of Eq.~\eqref{Htilde} with respect to the variational state:
\begin{equation}
\label{PsiChevy2}
	\vert \psi \rangle = g_{LM} \vert 0 \rangle \vert L M 0 \rangle + \sum_{k \lambda n}  \alpha_{k \lambda n}   \bed_{k \lambda n} \vert 0 \rangle \vert L M n \rangle
\end{equation}
Minimization with respect to $\alpha^*_{k\lambda n}$ and $g_{LM}^*$ yields the following equations:
\begin{equation}\label{varequation1}
\left[-E+BL(L+1)\right]g_{LM} + B\sqrt{L(L+1)}\sum_{k\lambda}\xi_{k\lambda}\alpha_{k\lambda n}=0
\end{equation}
and 
\begin{eqnarray}\label{varequation2}
&&\left[-E+BL(L+1)+W_{k\lambda}\right]\alpha_{k\lambda n}-2 B \sum_{\nu} \boldsymbol{\sigma}^\lambda_{ n\nu} \boldsymbol{\eta}^L_{ n\nu}\alpha_{k\lambda\nu}\nonumber\\
&&+B\delta_{ n,\pm1}\xi_{k\lambda}\sum_{k'\lambda'}\xi_{k'\lambda'}\alpha_{k'\lambda' n}\nonumber\\
&&=-B\sqrt{L(L+1)}\xi_{k\lambda}g_{LM}\delta_{ n,\pm1}
\end{eqnarray}
where we defined $\delta_{ n,\pm 1}=\delta_{ n,1}+\delta_{ n,-1}$,  $\xi_{k\lambda}=\sqrt{\lambda(\lambda+1)}V_\lambda(k)/W_{k\lambda}$, $W_{k\lambda}=\omega_k + B \lambda (\lambda+1)$, and $ \boldsymbol{\eta}^L_{ n\nu}=\bra{LM n}\hat {\vec J}'\ket{LM\nu}$. In what follows,  we show that Eqs. \eqref{varequation1} and \eqref{varequation2} can be solved in closed form. 

First, the angular-momentum coupling term of Eq.~(\ref{varequation2}) is given by:
\begin{eqnarray}
 \boldsymbol{\sigma}^\lambda_{ n\nu} \boldsymbol{\eta}^L_{ n\nu}&=& n^2\delta_{ n\nu}+\frac{1}{2}\sqrt{\lambda (\lambda+1)-\nu(\nu+1)}\nonumber\\
&\times&\sqrt{L (L+1)-\nu(\nu+1)}\delta_{ n,\nu+1}\nonumber\\
&+&\frac{1}{2}\sqrt{\lambda (\lambda+1)-\nu(\nu-1)}\nonumber\\
&\times&\sqrt{L (L+1)-\nu(\nu-1)}\delta_{ n,\nu-1}
\end{eqnarray}
Assuming that $V_\lambda(k)\neq 0 $ for $\lambda=0,1$ only, we obtain that Eqs.~\eqref{varequation1} and~\eqref{varequation2} are solved by $\alpha_{k\lambda n}=0$ for $\lambda  = 0$. Consequently,  $g_{LM}$, $\alpha_{k1\pm1}$, and $\alpha_{k 1 0}$ are the only variational parameters.\\

For   $\alpha_{k 1 0}$ we obtain
\begin{eqnarray}\label{alphaeq1}
&&\left[-E+BL(L+1)+\omega_k+2B\right]\alpha_{k10}\nonumber\\&& - B\sqrt{2L (L+1) }(\alpha_{k11}+\alpha_{k1-1})=0
\end{eqnarray}

For the $\alpha_{k1\pm1}$ components   we find two identical equations
\begin{eqnarray}\label{alphaeq2}
&&\left[-E+BL(L+1)+\omega_k\right]\alpha_{k1,\pm1}\nonumber\\&& - B\sqrt{2L (L+1) }\alpha_{k10}+B\xi_{k1}\sum_{k'}\xi_{k' 1}\alpha_{k'1,\pm1}\nonumber\\
&=&-B\sqrt{L(L+1)} \xi_{k1} g_{LM}
\end{eqnarray}
By symmetry we expect $|\alpha_{k11}|=|\alpha_{k1-1}|$, however, if $\alpha_{k11}=-\alpha_{k1-1}$ were true, Eq.~\eqref{alphaeq1} would imply $\alpha_{k10}=0$. This in turn would lead to a contradiction in Eq.~\eqref{alphaeq2} which shows that $\alpha_{k11}=\alpha_{k1-1}$.

Thus, from Eq.~\eqref{alphaeq1} we obtain 
\begin{equation}
\alpha_{k10}=\frac{2B \sqrt{2L(L+1)}}{-E+\omega_k +B L (L+1) +2B} \alpha_{k11}
\end{equation}

Let us now define the inverse propagator
\begin{equation}
P_E(k)=BL(L+1)-E + \omega_k - \frac{4B^2 L(L+1)}{-E+\omega_k +B L (L+1) +2B}
\end{equation}
and rewrite Eq.~\eqref{alphaeq2} as:
\begin{equation}\label{alphaxx}
\alpha_{k11}=-\frac{B \xi_{k 1}}{P_E(k)} \sum_{k'}\xi_{k' 1}\alpha_{k' 1 1}-\frac{B\sqrt{L(L+1)}\xi_{k1}}{P_E(k)}g_{LM}
\end{equation}
 In addition it is convenient to introduce the  variable $\chi$ as
\begin{equation}
g_{LM}\chi=\sum_{k}\xi_{k 1}\alpha_{k11}
\end{equation}
After multiplying Eq.~\eqref{alphaxx} with $\xi_{k1}$ and  integration over $k$ we find
\begin{equation}
\chi=-B\sqrt{L(L+1)}\frac{\int_0^\infty dk\, \xi_{k1}^2/P_E(k)}{1+B \int_0^\infty dk\, \xi_{k1}^2/P_E(k)}
\end{equation}\\

This finally yields the Dyson equation
\begin{equation}
-E+BL(L+1)-\Sigma_L(E)=0
\end{equation}
where the self-energy is given by
\begin{equation}\label{selfenergyapp}
\Sigma_L(E)=B^2 L (L+1) \frac{\int_0^\infty dk\, \xi_{k1}^2/P_E(k)}{1+B \int_0^\infty dk\, \xi_{k1}^2/P_E(k)}
\end{equation}
and
\begin{equation}
\xi_{k1}=\sqrt{2}\frac{V_1(k)}{\omega_k+2B}
\end{equation}
Furthermore, we absorbed the deformation energy $E_\text{def}$, Eq.~(\ref{defEnergy}), which is identical for all the $L$-levels, into the definition of $E$.  Note, that if $B/u_\lambda \gg 1$, the self energy   $\Sigma_L\to BL(L+1)$ and the Dyson equation is solved by $E=0$. This means that for weak interactions the impurity levels are shifted by the mean-field deformation energy only. \\

The self-energy of Eq.~\eqref{selfenergyapp} can be partially evaluated analytically. It is convenient to define
\begin{equation}
\omega=E-BL(L+1)
\end{equation}
and to rewrite the retarded self-energy, $\Sigma_L^\text{ret}(\omega) \equiv \Sigma_L(\omega+i0^+)$, as
\begin{equation}
\Sigma_L^{\text{ret}}(\omega)=2 B^2 L (L+1) \frac{\chi_L(\omega)}{1+2 \chi_L(\omega)}
\end{equation}
where
\begin{equation}\label{eqchi}
\chi_L(\omega)=\int_0^\infty dk\, \frac{V_1(k)^2}{[\omega_k+2B]^2}\frac{1}{P_{\omega+i 0^+}(k)}.
\end{equation}
The integrand of $\chi_L(\omega)$ possesses poles   at the momenta $k_0$ satisfying  $\omega_{k_0}=\omega$ for $L=0$ and at the momenta $k_{1,2}$ satisfying $\omega_{k_{1}}=\omega+2 B L $ and $\omega_{k_{2}}= \omega-2B L(L+1) $ for states with $L>0$. Using the relation $1/(x+i0^+)=\mathcal{P}(1/x)-i\pi\delta(x)$ this reveals the onset of the scattering continua in the spectral function. 

For $L=0$ one finds
\begin{equation}
\text{Im}\chi_{L=0}(\omega)=\pi\theta(\omega)\zeta_0
\end{equation}
where
\begin{equation}
\zeta_0=\frac{V_1(k_0)^2}{[\omega+2 B]^2}\left.\left[\frac{\partial \omega_k}{\partial k}\right]^{-1}\right|_{k=k_0}
\end{equation}
while for $L>0$ one has
\begin{eqnarray}
\text{Im}\chi_{L>0}(\omega)&=&\frac{\pi}{2}\theta(\omega_{k1})\left[1-\frac{1}{\sqrt{1+4 L (L+1)}}\right]\zeta_1\nonumber\\&+&
\frac{\pi}{2}\theta(\omega_{k2})\left[1+\frac{1}{\sqrt{1+4 L (L+1)}}\right]\zeta_2\nonumber\\
\end{eqnarray}
where
\begin{equation}
\zeta_{1,2}=\frac{V_1(k_0)^2}{[\omega+2 B]^2}\left.\left[\frac{\partial \omega_k}{\partial k}\right]^{-1}\right|_{k=k_{1,2}}
\end{equation}
Finally, the real part of  $\chi_L (\omega)$ follows from the  principal value integration.

\section{Deformation of the phonon density}
\label{sec:deformation}

From Eq.~(\ref{ArArlm}) we obtain the expression for the phonon density in the rotating impurity frame:
\begin{multline}
\label{PhDens}
	n(\mathbf{r}) \equiv \langle \bd_\mathbf{r} \be_\mathbf{r} \rangle \\= \frac{1}{r^2} \sum_{\underset{\lambda' \mu'}{\lambda \mu} } ~i^{-\lambda+\lambda'}~Y_{\lambda \mu} (\Theta_r,\Phi_r) Y^\ast_{\lambda' \mu'} (\Theta_r,\Phi_r)  \langle \bd_{r\lambda \mu} \be_{r\lambda' \mu'} \rangle
\end{multline}

Using Eq. (\ref{brlmViabklm}), we evaluate the partial-wave contributions:
\begin{multline}
\label{brlmViabklmAver}
	\langle \be^\dagger_{r \lambda \mu}  \be_{r \lambda' \mu'} \rangle \\= i^{\lambda-\lambda'} \frac{2}{\pi} r^2 \int k dk \int k' dk'~j_\lambda (kr) j_{\lambda'} (k'r)~\langle \be^\dagger_{k\lambda\mu} \be_{k'\lambda' \mu'} \rangle
\end{multline}

We calculate the expectation values, $\langle \dots \rangle$, with respect to the states in the transformed frame, $\vert \phi \rangle = \hat U \vert \psi \rangle$, where $\hat U$ and  $\vert \psi \rangle$ are given by Eqs. (\ref{Utransf}) and (\ref{PsiChevy}) of the main text. Finally, the expectation values of $\langle \be^\dagger_{k\lambda\mu} \be_{k'\lambda' \mu'} \rangle$ are given by:
\begin{multline}
\label{blkmAver}
	\langle \be^\dagger_{k\lambda\mu} \be_{k'\lambda' \mu'} \rangle = \delta_{\mu 0} \delta_{\mu' 0} \Biggl [ 3 \frac{V_\lambda(k)}{W_{k \lambda}} \frac{V_{\lambda'} (k')}{W_{k' \lambda'}}  -  g^{\ast}_{LM} \alpha_{k' \lambda' 0} \frac{V_\lambda(k)}{W_{k \lambda}} \\- g_{LM} \alpha^\ast_{k \lambda 0} \frac{V_{\lambda'} (k')}{W_{k' \lambda'}}\frac{V_\lambda(k)}{W_{k \lambda}}  + \vert \alpha_{k \lambda \mu}\vert^2 \Biggr ]
\end{multline}

\clearpage


%

\end{document}